\documentclass[preprint,showpacs,preprintnumbers,amsmath,amssymb,prb,superscrip
taddress
]{revtex4}
\usepackage{graphicx}
\usepackage{dcolumn}
\usepackage{bm}

\begin{document}

\title{Structure, magnetic and transport properties of Ti-substituted 
La$_{0.7}$Sr$_{0.3}$MnO$_{3}$}

\author{M. S. Kim}
\affiliation{Department of Physics, University of Missouri-Rolla, Rolla,
MO 65409}
\affiliation{Graduate Center for Materials Research, University of 
Missouri-Rolla, Rolla, MO 65409}
\author{J. B. Yang}
\affiliation{Department of Physics, University of Missouri-Rolla, Rolla,
MO 65409}
\affiliation{Graduate Center for Materials Research, University of 
Missouri-Rolla, Rolla, MO 65409}
\author {Q. Cai}
\affiliation{Department of Physics, University of Missouri-Columbia, Columbia,
MO 65211}
\author{X. D. Zhou}
\affiliation{Graduate Center for Materials Research, University of 
Missouri-Rolla, Rolla, MO 65409}
\author{W. J. James}
\affiliation{Graduate Center for Materials Research, University of 
Missouri-Rolla, Rolla, MO 65409}
\affiliation{Department of Chemistry, University of Missouri-Rolla, Rolla,
MO 65409} 
\author{W. B. Yelon}
\affiliation{Graduate Center for Materials Research, University of 
Missouri-Rolla, Rolla, MO 65409}
\affiliation{Department of Chemistry, University of Missouri-Rolla, Rolla,
MO 65409}
\author{P. E. Parris}
\affiliation{Department of Physics, University of Missouri-Rolla, Rolla,
MO 65409}
\author{D. Buddhikot}
\affiliation{Tata Institute of Fundamental Research, Colaba, Mumbai 400-005, 
India}
\author{S. K. Malik}
\affiliation{Tata Institute of Fundamental Research, Colaba, Mumbai 400-005, 
India}

\begin{abstract}

Ti-substituted perovskites, La$_{0.7}$Sr$_{0.3}$Mn$_{1-x}$Ti$_{x}$O$_{3}$,
with 0$\leq$ x $\leq$ 0.20, were investigated by neutron diffraction, 
magnetization, electric resistivity, and magnetoresistance(MR) measurements. 
All samples show a rhombohedral structure (space group $R\overline{3}c$) from 
10 K to
room temperature. At room temperature, the cell parameters $a$, $c$  and the 
unit cell
 volume increase with increasing Ti content. However, at 10 K, the cell 
parameter $a$ 
has a maximum value for 
$x=0.10$, and decreases for $x>0.10$, while the unit cell volume remains nearly
constant for $x>0.10$. The average (Mn,Ti)-O bond length increases up to 
x=0.15, and 
the (Mn,Ti)-O-(Mn,Ti) bond angle decreases with increasing Ti content to its 
minimum 
value at x=0.15 at room temperature. Below the Curie temperature T$_{C}$, the 
resistance exhibits metallic behavior for the $x\leq 0.05$ samples. A 
metal(semiconductor) to insulator transition is observed for the $x\geq 0.10$ 
samples. 
A peak in resistivity appears below T$_{C}$ for all samples, and shifts to a 
lower 
temperature as x increases. The substitution of Mn by Ti decreases the $2p-3d$ 
hybridization between O and Mn ions, reduces the bandwidth $W$, and increases 
the 
electron-phonon coupling. Therefore, the T$_C$ shifts to a lower temperature 
and the 
resistivity increases with increasing Ti content. A field-induced shift of the 
resistivity maximum occurs at $x$ $\leq$ 0.10. The maximum MR effect is about 
70$\%$ for La$_{0.7}$Sr$_{0.3}$Mn$_{0.8}$Ti$_{0.2}$O$_{3}$. The separation of 
T$_C$ and 
the resistivity maximum temperature T$_{\rho,max}$ enhances the MR effect in 
these 
compounds due to the weak coupling between the magnetic ordering and the 
resistivity as 
compared with La$_{0.7}$Sr$_{0.3}$MnO$_{3}$.  
\end{abstract}

\maketitle

\section{Introduction}

The A$_{1-x}$A'$_x$MnO$_3$ perovskites are interesting systems because of the 
anomalous 
magnetic and transport properties exhibited by them such as colossal 
magnetoresistance 
(CMR), metal-insulator transitions, antiferromagnetic-ferromagnetic ordering, 
and 
lattice dynamics associated with phase transitions. 
\cite{searle,goodenough,kusters,helmolt,jin,unishibara,tomioka,tokura} Zener's 
double 
exchange (DE) interaction between Mn$^{3+}$ and Mn$^{4+}$ ions through charge 
carriers 
in the oxygen $2p$ orbitals was introduced in order to explain the coupling of 
magnetic 
and electronic properties in these compounds.\cite{zener,anderson,gennes,kubo} 
Undoped 
LaMnO$_3$ is an \emph{A}-type antiferromagnetic insulator. By substitution of 
La$^{3+}$ 
with a divalent cation, LaMnO$_3$ can be driven into a metallic and 
ferromagnetic 
state.  Both Mn$^{3+}$ and Mn$^{4+}$ ions possess a local spin (S=3/2) from 
their lower 
$t_{2g}^3$ orbitals, and Mn$^{3+}$ has an extra electron in the $e_g$ orbital 
which is 
responsible for conduction. The spin of the $e_g^1$ electron in Mn$^{3+}$ is 
ferromagnetically coupled to the local spin of $t_{2g}^3$ according to Hund's 
rule. Sr 
doping induces holes in the $e_g$ band near the Fermi energy, producing mobile 
holes 
and conduction. However, recent studies have shown that DE is not sufficient to 
explain 
the complex physics in these compounds, especially as regards  the lattice 
distortions 
coinciding with the emergence of CMR \cite{millis,millis2}. An understanding of 
the 
Sr-doped systems requires one to consider both DE interactions in the 
Mn$^{3+}$-O-Mn$^{4+}$ pairs and the strong electron-phonon coupling, including 
lattice 
polarons and dynamic Jahn-Teller(J-T) distortions.\cite{dai,radaelli} The 
polaron 
effect arising from J-T distortion was introduced to explain the electronic 
transport 
mechanism in the high temperature region, T $\approx$ T$_{C}$, where a strong 
electron-phonon interaction is required to reduce the kinetic energy of the 
conduction 
electrons. The local J-T distortion of the MnO$_{6}$ octahedron lowers the 
energy of 
the $e^1_{g}$ electron and the charge carrier can then be localized to form a 
lattice 
phonon. Therefore local lattice distortion above T$_{C}$ rapidly decreases 
electron 
hopping, thus increasing the resistivity.\cite{millis2} Recently, it was found 
that 
polaron hopping was also the dominant conduction mechanism below 
T$_{C}$.\cite{ju,dessau,kim,zhao} A sharp increase of polaron density at 
temperatures 
below T$_{C}$ leads to a charge carrier density collapse, which is related to 
the 
resistivity peak and the CMR of doped manganites.\cite{alexandrov}

In order to understand the unusual magnetic and transport properties of doped 
perovskites A$_{1-x}$D$_{x}$MnO$_{3}$, many studies have been carried out by 
doping the 
trivalent rare earth site(A site) with divalent atoms(Ca, Sr, Ba, 
etc).\cite{jin,neumeier,muroi,medarde,cao,xiaming} 
Other studies have also shown that substitution for Mn (B site) dramatically 
affects 
the magnetic and transport properties of perovskites.\cite{martin,ahn,blasco}
The B site modification has merit in that it directly affects the Mn network by 
changing the Mn$^{3+}$/Mn$^{4+}$ ratio and the electron carrier density.  
Therefore in 
order to better understand the role of Mn and its local environment in 
La$_{0.7}$Sr$_{0.3}$MnO$_{3}$, we studied the effects of replacing some of the 
Mn with 
Ti. The structural, magnetic and electrical phase transitions and transport 
properties 
of La$_{0.7}$Sr$_{0.3}$Mn$_{1-x}$Ti$_{x}$O$_{3}$ with 0 $\leq$x$\leq$ 0.20 have 
been 
investigated by neutron diffraction, magnetization, electric resistivity, and 
magnetoresistance measurements and the results are presented here.

\section{Experimental}
Samples of Ti-substituted La$_{0.7}$Sr$_{0.3}$Mn$_{1-x}$Ti$_{x}$O$_{3}$, with 
$0\leq 
x\leq 0.2$, were prepared using the conventional solid state reaction method. 
Highly
purified La$_{2}$O$_{3}$, SrCO$_{3}$, TiO$_{2}$, MnO were mixed in 
stoichiometric 
ratios, ground, and then pelletized under 10,000 psi pressure to a 1 cm 
diameter. The 
pelletized samples were fired at $1500^{\circ}C$ in air for 12 hours, then 
reground and 
sintered at $1250^{\circ}C$ for 24 hours in air. X-ray diffraction of the 
powders was 
carried out at room temperature using a SCINTAG diffractometer with 
Cu-K$\alpha$
radiation. X-ray diffraction data indicated all samples to be single phase. 
Powder
neutron diffraction experiments were performed at the University of 
Missouri-Columbia 
Research Reactor(MURR) using neutrons of wavelength $\lambda$=1.4875 $\AA$. The 
data 
for each sample were collected between 2$\theta$ = 5.65 - 105.60$^{\circ}$ at 
300K and 
10 K. Refinement of the neutron diffraction data was carried out using the 
FULLPROF 
program\cite{full}, which permits multiple phase refinements as well as 
magnetic 
structure refinements. Magnetic measurements were conducted with a SQUID 
magnetometer(MPMS, Quantum design). The magnetization curves with zero-field 
cooling(ZFC) and field cooling(FC) were measured in an applied magnetic field 
of 50Oe.
Resistivity data were obtained using a physical properties measurement 
system(PPMS, 
Quantum design) with a standard four-point probe method.

\section{Results and Discussion}

Figure 1 shows the x-ray diffraction patterns of 
La$_{0.7}$Sr$_{0.3}$Mn$_{1-x}$Ti$_{x}$O$_{3}$ samples, with $0\leq 
x\leq 0.2$, at room temperature(RT). All the samples are single phase and all 
peak 
positions can be indexed to La$_{0.67}$Sr$_{0.33}$MnO$_{2.91}$(JCPDS 50-0308), 
space 
group $R$$\overline{3}$c. In order to investigate the details of the structural 
distortion and the magnetic interactions in these compounds, powder neutron 
diffraction 
measurements were performed at different temperatures. Figure 2 shows the 
neutron 
diffraction patterns of La$_{0.7}$Sr$_{0.3}$Mn$_{1-x}$Ti$_{x}$O$_{3}$ with 
$x=0.0$, x=0.03 and 
0.15 measured at RT and 10 K. All patterns can be fitted with the 
$R$$\overline{3}$c 
rhombohedral space group (No. 167) in which the atomic positions are La(Sr): 6a 
(0,0,1/4), Mn(Ti): 6b (0,0,0); O18e (x,0,1/4). The $P1$ space group was used to 
fit the 
magnetic structure with collinear Mn magnetic moments because of its 
flexibility.   Refined structural and magnetic parameters are listed in Tables 
I and II for RT and 10K, respectively. For samples with $x \geq 0.10$, there is 
no magnetic ordering at RT since T$_{C}<$ RT, whereas for 
samples with $x \leq 0.10$, T$_{C}>$ RT magnetic ordering is observed. The 
arrows on 
the neutron diffraction patterns of the $x=0.0$ sample (Figure 2 ) indicate 
magnetic 
reflection peaks that are not present for the $x=0.20$ sample at RT. The peak 
intensities of the magnetic reflections decrease with Ti substitution at both 
RT and 10 
K. In addition, the refinement results confirm that the substituted Ti ions go 
into B 
sites, not into A sites, because the ionic radius of Ti$^{4+}$ (0.605$\AA$) 
lies 
between the ionic radius of Mn$^{4+}$ (0.530$\AA$) and Mn$^{3+}$ 
(0.645$\AA$)\cite{shannon}. The tolerance factor, which is the geometric 
measure of 
size mismatch of perovskites, 

\begin{equation}
	t=(r_{(La,Sr)}+r_{O})/[(r_{(Mn,Ti)}+r_{O})\sqrt{2}]
\end{equation}
 
decreases linearly from 0.928 for La$_{0.7}$Sr$_{0.3}$MnO$_{3}$ to 0.921 for 
La$_{0.7}$Sr$_{0.3}$Mn$_{0.8}$Ti$_{0.2}$O$_{3}$, which is in the stable range 
of the 
perovskite structure 0.89$<$t$<$1.02.\cite{xiaming} Therefore, substitution of 
Mn by Ti 
does not change the crystal structure itself but changes the bond lengths and 
the bond 
angles of the MnO$_{6}$ octahedra.

Fig. 3 plots the lattice parameters $a$,$c$ and the unit cell volumes  of 
La$_{0.7}$Sr$_{0.3}$Mn$_{1-x}$Ti$_{x}$O$_{3}$ versus the Ti(Ti$^{4+}$) content 
at room temperature and 10 K. The lattice parameters $a$,$c$ and the unit cell 
volume increase with the Ti content at 
RT. At 10 K, the lattice parameter $a$ shows a maximum value at $x=0.10$ and 
then 
decreases as $x>0.10$, and the unit cell volume increases up to $x=0.10$ and 
becomes 
almost constant for $x>0.10$. The refined magnetic moment of the Mn atoms 
indicate that Mn atoms have a high spin state, and the 
average valence state of the Mn varies from 3d$^{3.5}$ to 3d$^{3.3}$ for x=0.0 
and 
x=0.15, which suggests that the Ti atoms are in the Ti$^{4+}$ state. The values 
of the temperature factor B of oxygen increase with increasing Ti content which 
is consistent with the increase of the Mn-O bond length. This is likely related 
to the structural disorder in the position of oxygen atoms due to the 
substitution of Mn for Ti.

The average (Mn,Ti)-O bond length and (Mn,Ti)-O-(Mn,Ti) bond angle extracted 
from the 
Rietveld refinement at RT and 10K are shown in Figure 4.  
The bond length of La$_{0.7}$Sr$_{0.3}$Mn$_{1-x}$Ti$_{x}$O$_{3}$ increases up 
to 
$x=0.15$ and remains constant thereafter for $x\geq 0.15$, while the bond angle 
decreases and attains an anomalous minimum value for $x=0.15$ at RT. 
At 10K, the bond length increases up to $x=0.10$ and remains constant for 
$x>0.10$, 
while the bond angle decreases with increasing $x$.
The bond length and the bond angle are closely related to the oxygen positions. 
Therefore, an increasing (Mn-Ti)-O bond length and a decreasing 
(Mn,Ti)-O-(Mn,Ti) bond 
angle are strongly correlated. The changes in bond length and bond angle of 
MnO$_{6}$ 
compensate one another to diminish the internal strain induced by Ti$^{4+}$.
Since the exchange interaction between Mn-Mn depends on both the bond angle and 
the 
bond distance, the decrease in bond angle and the increase in bond length 
decrease the 
Mn-Mn exchange interaction which leads to a lower magnetic ordering temperature 
T$_C$(see later discussion of M-T curves).

The electronic bandwidth $W$ has been used to discuss magnetic and transport 
properties 
of perovskites with varied A-site doping.\cite{radaelli,hwang}. The empirical 
formula 
of the bandwidth $W$ for $ABO_{3}$-type perovskites using the tight binding 
approximation\cite{medarde} is
\begin{equation} 
	W \propto \frac{cos\omega}{(d_{Mn-O})^{3.5}},
\end{equation} 
 where $\omega = \frac{1}{2} ( \pi -\langle Mn-O-Mn \rangle)$ and d$_{Mn-O}$ is 
the 
Mn-O bond length. The calculated values of ${cos\omega}/{(d_{Mn-O})^{3.5}}$ 
using the  
refinement results are shown in Figure 4(c). We assumed the bandwidth $W$ is 
proportional
to the values of ${cos\omega}/{(d_{Mn-O})^{3.5}}$. It is found that the 
bandwidth $W$ 
decreases with increasing Ti content. Further, the bandwidth at RT is smaller 
than the 
bandwidth at 10K for a given Ti content. The evolution of the bandwidth follows 
the 
change in the $\langle Mn-O-Mn \rangle$ bond angle. The decrease in bandwidth 
reduces 
the overlap between the O-$2p$ and the Mn-$3d$ orbitals, which in turn 
decreases the 
exchange coupling of Mn$^{3+}$-Mn$^{4+}$, and the magnetic ordering temperature 
T$_C$ 
as well. 
For a charge-transfer insulator, the band gap energy $E_{g}$ in the insulating 
phase 
can be written as $E_{g}=\Delta-W$, where $\Delta$ is the charge-transfer 
energy and 
$W$ is the O-$2p$-like bandwidth. 
In practice, $\Delta$ changes little in the La$_{1-x}$Sr$_{x}$MnO$_{3}$ system 
and thus 
the bandwidth $W$ becomes a main factor in tuning the band gap 
energy.\cite{harrison} 
For the La$_{0.7}$Sr$_{0.3}$Mn$_{1-x}$Ti$_{x}$O$_{3}$ compounds, the decrease 
in 
bandwidth $W$ increases the band gap, $E_{g}$, and leads to the metal to 
insulator 
transition for $x > 0.10$.

Figure 5 shows the magnetization versus temperature (M-T) curves measured under 
field-cooled (FC) and zero field-cooled (ZFC) conditions in a magnetic field of 
50 Oe 
for the $x=0.05, 0.10$, and 0.15 samples. A sharp paramagnetic to ferromagnetic 
transition is observed at a critical temperature T$_{C}$. Figure 6 shows the 
Curie 
temperatures, T$_{C}$, of La$_{0.7}$Sr$_{0.3}$Mn$_{1-x}$Ti$_{x}$O$_{3}$ for 
differing 
Ti content. The decrease in T$_{C}$ is obviously related to the changes in 
bandwidth as 
seen in Fig. 4(c). The T$_{C}$ drops at a rate of about 10K per Ti. 
$\lambda$-shaped 
magnetization curves in ZFC emerge for $x \geq 0.10$ samples. The existence of 
$\lambda$-shaped curves under ZFC may be evidence of the formation of 
ferromagnetic 
clusters with a spin glass state. The Ti substitution weakens the exchange 
interaction 
and breaks the Mn-O-Mn network, and creates short range ordered ferromagnetic 
clusters. 
As more Ti is substituted, more inhomogeneous clusters are formed, which leads 
to a 
broadening of the paramagnetic to ferromagnetic phase transition peak. A 
similar 
phenomenon has been observed in the La$_{0.7}$Ca$_{0.3}$Mn$_{1-x}$Ti$_x$O$_3$ 
system.\cite{xiaming}

Magnetization versus field (M-H) curves of 
La$_{0.7}$Sr$_{0.3}$Mn$_{1-x}$Ti$_{x}$O$_{3}$ at different temperatures are 
plotted in 
Figure 7.
At 20 K, all samples reach a nearly constant value of magnetization under a 
field, H = 
0.6 T. 
The estimated magnetic moments of the $x=$ 0.0, 0.05, 0.10 and 0.15 samples 
from 
magnetization data at 20 K are 3.79, 3.54, 3.24, and 2.49 $\mu_{B}$ per Mn 
atom, 
respectively. These moment values are in good agreement with the neutron 
diffraction 
refinement results (see Table II). The theoretically estimated magnetic moments 
of Mn 
from its valence state taking into account the dilution effect of Ti$^{4+}$, 
are 3.70, 3.55, 
3.40 and 3.35 $\mu_{B}$ respectively. This suggests that the decrease of  
magnetization 
with increasing Ti content is not only due to the dilution of magnetic 
Mn$^{4+}$ atoms 
but also due to the weakening of exchange coupling by the cluster formation.

Figure 8 shows the resistivity as a function of temperature under different 
applied 
fields for La$_{0.7}$Sr$_{0.3}$Mn$_{1-x}$Ti$_{x}$O$_{3}$ compounds with $x 
=0.0$, 
0.05, 0.10 and x=0.15. In the temperature range 4 - 300 K, the resistivity of 
the samples increases as the Ti content increases. The resistivity of 
0.05$\leq$ $x\leq $ 0.10 shows a maximum value at temperature 
T$_{\rho,max}$ below T$_{C}$, and then decreases as the temperature decreases. 
Finally 
the resistivity increases again as T decreases further for $x\geq 0.10$. The 
difference 
between T$_{C}$ and T$_{\rho, max}$ becomes larger as the Ti content increases 
and 
T$_{\rho,max}$ is lower than T$_{C}$. This behavior is quite different from 
that 
observed in the Ti-substituted La$_{0.7}$Ca$_{0.3}$Mn$_{1-x}$Ti$_{x}$O$_{3}$ 
series 
which exhibit large differences between T$_{C}$ and T$_{\rho,max}$\cite{cao}, 
and 
T$_{\rho,max}$ is higher than T$_{C}$.\cite{xiaming}.
For the $x \leq 0.15$ sample, a metal(semiconductor) to insulator transition 
(MIT) 
appears in the low temperature region. The field-induced shift of maximum 
resistivity 
to higher temperature appears for the $x \leq 0.10$ samples, and becomes 
negligible for 
$x \geq 0.15$. 
The suppression of the resistivity by the applied magnetic field occurs over 
the entire 
temperature range for all samples. At T $>$ T$_{C}$, the suppression of the 
resistivity 
becomes weaker. According to the DE mechanism, the mobility of the charge 
carriers 
$e_{g}$ electrons improves if the localized spins are polarized. The applied 
field 
aligns the canted electron spins which should reduce the scattering of 
itinerant 
electrons with spins and thus the resistivity is reduced. Therefore an applied 
magnetic 
field competes with the thermal fluctuations and maintains magnetic ordering 
around  
T$_{C}$ for the $x\leq 0.10$ samples, and thus shifts the T$_{\rho,max}$ to 
higher 
temperatures. 

Figure 9 shows the typical temperature dependence of the magnetoresistance [MR 
= 
$(\rho_{0}-\rho_{H}) / \rho_{0} \times 100$] of 
La$_{0.7}$Sr$_{0.3}$Mn$_{1-x}$Ti$_{x}$O$_{3}$ samples with 0$\leq$ x $\leq$ 
0.20 under an 
applied field of 1 and 3T. The maximum magnetoresistance increases with 
increasing Ti concentration. For example, the maximum MR values are 30$\%$, 
55$\%$ under 3T for 
$x=0.05$, $0.15$, respectively. The temperature of the MR peak shifts to a 
lower 
temperature, approximately 15 K per Ti. It is known that in A-site, 
electron-doped 
A$_{1-x}$A$'$$_x$MnO$_3$(x=0.3) compounds, the metal-insulator transition 
temperature 
T$_{MI}$ coincides with the T$_c$, and the metal-insulator transition is 
strongly 
coupled with the magnetic ordering transition. Therefore, a strong variation of 
the 
electrical resistivity up to several orders of magnitude, namely the colossal 
magnetoresistance (CMR) effect, occurs upon application of a magnetic field 
near 
T$_{C}$. However, for La$_{0.7}$Sr$_{0.3}$Mn$_{1-x}$Ti$_{x}$O$_{3}$ compounds, 
the 
T$_{C}$ is different from the metal to insulator transition temperature (MIT). 
The 
application of a magnetic field has much more effect on the change of electric 
resistivity when compared to La$_{0.7}$Sr$_{0.3}$MnO$_{3}$ due to the weak 
coupling 
between the MIT and the magnetic ordering. An enhancement of the MR effect is 
observed 
in these compounds, similar to that in 
La$_{0.7}$Ca$_{0.3}$Mn$_{1-x}$Ti$_{x}$O$_{3}$\cite{cao,xiaming} and 
Pr$_{1-x}$(Ca,Sr)$_{x}$MnO$_{3}$ compounds.\cite{maignan}
	
	The change of the electronic properties of Ti-substituted 
La$_{0.7}$Ca$_{0.3}$Mn$_{1-x}$Ti$_{x}$O$_{3}$ compounds is strongly related to 
the 
electron phonon coupling \cite{millis2}. Accordingly, in the 
La$_{1-x}$Sr$_{x}$MnO$_{3}$ system, the strong electron-phonon coupling 
localizes the 
conduction band electron as a polaron, due to the competition between the self 
trapping 
energy E$_{J,T}$ and the electron itinerant energy. The electron-phonon 
coupling 
constant $\lambda$ = E$_{J,T}$/$t$, where $t$ is the electron hopping parameter 
which is proportional to the electronic bandwidth $W$. As mentioned above, the 
substitution of 
Mn by Ti decreases the overlap of the O-$2p$ and Mn-$3d$ orbitals due to the 
decrease 
in $W$, thus increasing the electron-phonon coupling. This results in a shift 
of 
T$_{C}$ to lower temperatures and an increase of resistivity with increasing Ti 
content. As a consequence, one should consider a possible dependence of 
E$_{J,T}$ on Ti 
content. We cannot rule out the contribution from E$_{J,T}$, even though our 
data 
indicate that all the observed T$_{C}$ and resistivity changes can be 
explained, at 
least qualitatively, by the change in $W$. Especially, for $x \geq 0.15$, the 
electron-phonon coupling becomes very strong, and the insulator behavior occurs 
below 
T$_{C}$ as shown in Figure 8. The changes in bandwidth $W$ are not large enough 
to 
account for the dramatic changes in resistivity, and therefore, E$_{J,T}$ might 
be 
contributing significantly to the change of resistivity in these samples.        

	It has been proposed that, above T$_{C}$, charge may be localized in the 
form of 
J-T polarons.\cite{alexandrov} At T $\geq$ T$_{C}$, the resistivity of the CMR 
materials can be explained by the activated adiabatic polaron equation 
\cite{emin}

 \begin{equation}
	\rho=AT \exp (E_{hop}/kT)
\end{equation} 

 Figure 10 shows the plot of ln($\rho$/T) as a function 
of 1/T for La$_{0.7}$Sr$_{0.3}$Mn$_{1-x}$Ti$_{x}$O$_{3}$ compounds with 
$x=$0.00, 0.05, 0.10, 0.15 and 0.20 in the high temperature region with zero 
field resistivity data. Resistivity of all the samples 
shows a similar slope at T $\geq$ T$_{C}$, which can be fitted well with the 
small 
polaron model indicating the formation of a polaron. The polaron hopping energy 
$E_{hop}$ is calculated from the slopes. The calculated values of $E_{hop}$ are 
14.5 49.8, 
132.0, 138.3, and 152.5 $meV$ for $x=0.00, 0.05, 0.10, 0.15$, and 0.20, 
respectively. The increase of $E_{hop}$ is due to the substitution of Mn by Ti 
which depletes the oxygen $p$ holes and leads to an increase in the polaron 
binding energy. This further confirms 
that Ti substitution at Mn enhances the electron-phonon interaction, which 
decreases 
$W$ and increases $E_{hop}$ at high temperatures. The calculated polaron 
hopping energy 
shows a large variation between the $x=0.05$ and the $x\geq 0.10$ samples. This 
is in 
good agreement with the sharp increase in resistivity and its temperature 
dependence. 
Some studies have also suggested that polaron hopping is the prevalent 
mechanism to 
explain resistivity below T$_{C}$.\cite{hundley,pickett}. However, we were 
unable to 
fit the resistivity data of the Ti-doped La$_{0.7}$Sr$_{0.3}$MnO$_{3}$ samples 
below 
T$_{C}$ with several other polaron models such as the semiconducting 
model,\cite{mott} 
the variable range hopping(VHR) polaron model,\cite{sun} and the adiabatic 
polaron 
hopping model. Only the VHR polaron model works reasonably well for the low 
temperature 
region, T $<$ 75K, for the insulating state, $x=0.10$ sample. There may be 
other 
contributions, such as ferromagnetic clusters, which would increase the 
resistivity of 
the compound.

\section{Summary}

	The magnetic and electronic transport properties of Ti-substituted 
La$_{0.7}$Sr$_{0.3}$Mn$_{1-x}$Ti$_{x}$O$_{3}$ have been systematically 
investigated. 
All the Ti-substituted La$_{0.7}$Sr$_{0.3}$Mn$_{1-x}$Ti$_{x}$O$_{3}$ 
compositions have 
a rhombohedral structure, (space group $R\overline{3}c$). The correlation 
between 
ferromagnetic T$_{C}$ and T$_{\rho,max}$ becomes weaker and spin glass clusters 
are 
expected in the low temperature region with increasing Ti substitution. The 
resistivity 
in the high temperature region suggests the formation of localized polarons 
that affect 
the strong correlation between local structural changes and the MIT. The 
decrease of the 
bandwidth $W$ decreases the overlap between the O-$2p$ and Mn-$3d$ orbitals, 
which in 
turn decreases the exchange coupling of Mn-Mn and the magnetic ordering 
temperature 
T$_c$ as well. Our studies indicate that Ti substitution at Mn enhances the 
electron-phonon interaction in these compounds, which decreases the bandwidth 
and 
increases the resistivities in the entire temperature range. 

\begin{acknowledgements}
The authors thank Aranwela Hemantha for invaluable help in magnetoresistance 
measurements. The support by DOE under DOE contract \#DE-FC26-99FT400054 is 
acknowledged. 
\end{acknowledgements}

\newpage
\begin{table}
\caption{Refined parameters for 
La$_{0.7}$Sr$_{0.3}$Mn$_{1-x}$Ti$_{x}$O$_{3}$ compound with 
\emph{R$\overline{3}$c} 
space-group at room temperature(T=300K). Numbers in parentheses are statistical 
errors. 
$a$ 
and $c$ are the lattice parameters. $m$ is magnetic moment. $V$ is the unit 
cell 
volume. $B$ is the isotropic temperature parameter. 
$\chi^2$ is [R$_{wp}$/R$_{exp}$]$^2$ where R$_{wp}$ is the residual error of 
the 
weighted profile.}

\begin{ruledtabular}
\begin{tabular}{ccccccc}
Composition   & 0.00     & 0.03        & 0.05         & 0.10        & 0.15       
& 0.20 \\ 
\hline
$a$ ($\AA$)    & 5.5038(2) & 5.5107(1) & 5.5157(2) & 5.5225(2) & 5.5306(2) & 
5.5310(2) \\
$c$ ($\AA$)    & 13.3553(5) & 13.3635(4)& 13.3699(5)& 13.3845(5)& 13.4032(6) 
&13.4124(6) \\
$V$ ($\AA^{3}$) & 350.364(18) & 351.445(16) & 352.261(18) & 353.508(21) & 
355.042(22) &355.341(23)\\
\hline
$m$ ($\mu_{B}$) & 2.514(28) & 2.121(33)   & 1.022(63)   & 0.0  & 0.0   & 0.0  
\\
$\chi^{2}$ ($\%$) & 2.81  & 3.28    & 3.60    & 3.23   & 4.64  & 2.98 \\
\hline
O, 18e,x  & 0.5422(2)  & 0.5437(2) & 0.5448(2)  & 0.5460(2) & 0.5469(2) & 
0.5461(2)\\
B($\AA^{2}$), La(Sr),6a & 0.882(33)  & 0.8124(33)   & 0.873(35)    & 1.030(42)   
& 0.975(43)   & 1.149(40) \\
B($\AA^{2}$), Mn(Ti),6b & 0.423(54) & 0.556(56)   & 0.574(59)    & 0.364(66)   
& 0.394(67)   & 0.404(63) \\
B($\AA^{2}$), O,18e & 1.221(25)  & 1.248(25)   & 1.306(27)    & 1.501(37)   & 
1.475(37)   & 1.586(35)

\end{tabular}
\end{ruledtabular}
\end{table}

\newpage
\begin{table}
\caption{ Refined parameters for 
La$_{0.7}$Sr$_{0.3}$Mn$_{1-x}$Ti$_{x}$O$_{3}$ compound with 
\emph{R$\overline{3}$c} 
space-group at low temperature(T=10K). Numbers in parentheses are statistical 
errors. 
$a$ and $c$ are the lattice parameters. $m$ is magnetic moment. $V$ is the unit 
cell 
volume. $B$ is the isotropic temperature parameter. $\chi^2$ is 
[R$_{wp}$/R$_{exp}$]$^2$ where R$_{wp}$ is the residual error of the weighted 
profile.}

\begin{ruledtabular}
\begin{tabular}{ccccccc}
Composition  & 0.00   & 0.03        & 0.05         & 0.10        & 0.15       & 
0.20 \\ 
\hline
$a$ ($\AA$) & 5.4811(1) & 5.4940(1) & 5.4989(1) & 5.5116(1) & 5.5089(2) & 
5.5053(3) \\
$c$ ($\AA$) & 13.2756(3)    & 13.3037(4)& 13.3137(4)& 13.3354(4)& 13.3421(6) 
&13.3746(10) \\
$V$ ($\AA^{3}$) & 345.397(13) & 347.756(14) & 348.644(16) & 350.820(15) & 
350.652(22) &351.056(35)\\
\hline
$m$ ($\mu_{B}$) & 3.443(25) & 3.461(27)   & 3.506(32)   & 3.422(28)  & 
3.282(36)   & 2.913(51)  \\
$\chi^{2}$ ($\%$) & 3.23  & 3.51    & 2.89    & 2.69   & 3.72  & 4.90 \\
\hline
O, 18e,x  & 0.5431(1)  & 0.5442(2) & 0.5448(1)  & 0.5467(1) & 0.5469(2) & 
0.5472(2)\\
B($\AA^{2}$), La(Sr),6a & 0.167(26) & 0.240(28)   & 0.302(33)    & 0.318(28)   
& 0.449(37)   & 0.269(48) \\
B($\AA^{2}$), Mn(Ti),6b &  0.127(45) & 0.230(49)   & 0.271(57)    & 0.169(47)   
& 0.291(62)   & 0.174(81) \\
B($\AA^{2}$), O,18e & 0.328(21) & 0.536(23)   & 0.595(27)    & 0.649(22)   & 
0.926(28)   & 0.945(36)

\end{tabular}
\end{ruledtabular}
\end{table}

\newpage
\begin{figure}
\caption{X-ray diffraction patterns of 
La$_{0.7}$Sr$_{0.3}$Mn$_{1-x}$Ti$_{x}$O$_{3}$ 
(0$\leq$ x $\leq$ 0.20) at room temperature.}
\end{figure}

\begin{figure}
\caption{Neutron diffraction patterns of
La$_{0.7}$Sr$_{0.3}$Mn$_{1-x}$Ti$_{x}$O$_{3}$ (x=0.0, x=0.03 and x=0.20) at 10 
K and RT. (The
bottom curves(Yobs-Ycal) are the difference between experimental
data and refinement data. The vertical bars indicate the
magnetic(bottom) and Bragg(top) peak positions). Arrows indicate some of major 
magnetic diffraction peaks.}
\end{figure}

\begin{figure}
\caption{Lattice parameter $a$, $c$, and volume of 
La$_{0.7}$Sr$_{0.3}$Mn$_{1-x}$Ti$_{x}$O$_{3}$ versus Ti content at room 
temperature and at 10 K.}
\end{figure}

\begin{figure}
\caption{Average (Mn,Ti)-O bond lengths(a), (Mn,Ti)-O-(Mn,Ti) bond angles(b) 
and  
electronic bandwidth parameter ${cos\omega}/{(d_{Mn-O})^{3.5}}$(c), of 
La$_{0.7}$Sr$_{0.3}$Mn$_{1-x}$Ti$_{x}$O$_{3}$ at room temperature and at 10 K.}
\end{figure}

\begin{figure}
\caption {The magnetization versus temperature(M-T) curves of 
La$_{0.7}$Sr$_{0.3}$Mn$_{1-x}$Ti$_{x}$O$_{3}$ (x=0.05, 0.10, 0.15)  measured 
under 
field cooling (FC) and zero field cooling (ZFC) conditions in a magnetic field 
of 50 Oe 
.}
\end{figure}

\begin{figure}
\caption{The Curie temperature (T$_{C}$), and the temperature of maximum 
resistivity 
(T$_{\rho, max}$) of La$_{0.7}$Sr$_{0.3}$Mn$_{1-x}$Ti$_{x}$O$_{3}$ compounds 
with 
0$\leq$ x $\leq$ 0.20.}
\end{figure}

\begin{figure}
\caption{Field dependent magnetization of 
La$_{0.7}$Sr$_{0.3}$Mn$_{1-x}$Ti$_{x}$O$_{3}$ (0$\leq$ x $\leq$ 0.20) at 
different temperatures.} 
\end{figure}

\begin{figure}
\caption{Electric resistivity $\rho$ versus temperature for 
La$_{0.7}$Sr$_{0.3}$Mn$_{1-x}$Ti$_{x}$O$_{3}$ compounds (x=0.0(a), 0.05(b), 
0.10(c), and 
0.15(d)) in applied magnetic field H=0, 1, 3, and 5 T. Arrows indicate the 
T$_{\rho, 
max}$. The inset in (d) is the plot of  resistivity of x=0.15 compound (with 
log scale) 
in H= 0 T.}
\end{figure}

\begin{figure}
\caption{Temperature dependence of magnetoresistance of 
La$_{0.7}$Sr$_{0.3}$Mn$_{1-x}$Ti$_{x}$O$_{3}$ (0$\leq$ x $\leq$ 0.20) compounds 
in the 
magnetic field of H=1, 3 T.}
\end{figure}

\begin{figure}
\caption{ln($\rho$/T) versus 1/T plots in the high temperature region of 
La$_{0.7}$Sr$_{0.3}$Mn$_{1-x}$Ti$_{x}$O$_{3}$ (0$\leq$ x $\leq$ 0.20) compounds 
. 
Dot line is the fitting line .}
\end{figure}

\end{document}